\documentclass[3p,times,procedia]{elsarticle}
\usepackage{nupha_ecrc}

\volume{00}

\firstpage{1}

\journalname{Nuclear Physics A}

\runauth{}

\jid{nupha}

\jnltitlelogo{Nuclear Physics A}

\usepackage{amssymb}
\usepackage{color}

\usepackage[figuresright]{rotating}

\newcommand{\sNN}{s_\mathrm{NN}}

\begin{document}

\begin{frontmatter}


 \title{Vorticity in the QGP liquid and $\Lambda$ polarization at the RHIC Beam Energy Scan}

\dochead{}



\author[inst1,inst2]{Iurii Karpenko}
\author[inst1,inst3]{Francesco Becattini}

\address[inst1]{INFN - Sezione di Firenze, Via G. Sansone 1, I-50019 Sesto Fiorentino (FI), Italy}
\address[inst2]{Bogolyubov Institute for Theoretical Physics, 14-b, Ul.~Metrolohichna, 03680 Kiev, Ukraine}
\address[inst3]{Universit\'a di Firenze, Via G. Sansone 1, I-50019 Sesto Fiorentino (FI), Italy}

\begin{abstract}
Polarization of $\Lambda$ hyperons and their antiparticles is calculated in a 3+1 dimensional viscous hydrodynamic model with initial state from UrQMD hadron/string cascade. We find that, along with recent results from STAR, the mean polarization at fixed centrality decreases as a function of collision energy from 1.5\% at $\sqrt\sNN=7.7$~GeV to 0.2\% at $\sqrt\sNN=200$~GeV. We explore the effects which lead to such collision energy dependence, feed-down corrections and a difference between $\Lambda$ and $\bar\Lambda$.
\end{abstract}


\end{frontmatter}


\section{Introduction}\label{sec1}


There is growing interest recently, from both experimental and theoretical sides, to polarization of $\Lambda$ hyperons produced in relativistic heavy ion collisions. From the experimental side, there are recent measurements by STAR collaboration which for the first time show a significantly non-zero polarization of both $\Lambda$ and $\bar\Lambda$ produced in noncentral Au-Au collisions at the RHIC Beam Energy Scan (BES) program, $\sqrt\sNN=7.7\dots200$~GeV \cite{STAR:2017ckg}. From theory side, it has been proposed that a thermodynamic spin-vorticity coupling mechanism can result in spin polarization of hadrons produced out of vortical fluid. A derivation of this effect has been made in \cite{Becattini:2007nd,Becattini:2013fla}. This gave rise to a series of polarization calculations in hydrodynamic models \cite{Becattini:2015ska,Pang:2016igs,Karpenko:2016jyx,Xie:2017upb}.

In the present work we calculate the polarization of $\Lambda$ and anti-$\Lambda$ hyperons in Au-Au collisions at RHIC BES in a state-of-the-art hybrid model \texttt{UrQMD+vHLLE}, which is tuned to reproduce basic hadron observables for heavy ion collisions in the BES program: (pseudo)rapidity, transverse momentum distributions and elliptic flow coefficients.

\section{Model}

We start with a brief description of the hydro+cascade model used in the studies, whereas through description can be found it \cite{Karpenko:2015xea}.

For hydrodynamic modeling at BES energies it is essential to start with an initial state which is not boost invariant, and in addition to energy and momentum densities contains nonzero densities of conserved quantum numbers: baryon, electric charge and strangeness. We use UrQMD string/hadron cascade \cite{Bass:1998ca, Bleicher:1999xi} for this purpose. The initial state dynamics from UrQMD does not lead to thermalization, therefore a forced fluidization is performed at a hypersurface $\tau=\sqrt{t^2-z^2}=\tau_0$: energies and momenta of hadrons which cross the hypersurface, are distributed to neighboring fluid cells according to a 3 dimensional Gaussian profile. The minimal value of starting time $\tau_0$ is the time for the two colliding nuclei to completely pass through each other. This initial state has generally nonzero flow velocity components, especially in case of fluctuating initial state taken from a single UrQMD simulation. As a result, all components of thermal vorticity tensor are initially non-vanishing.

With such initial state, we run event-by-event 3 dimensional viscous hydrodynamic evolution with the help of \texttt{vHLLE} code \cite{Karpenko:2013wva}. The fluid to particle transition, or ``particlization'', is set to happen at a hypersurface of constant energy density (in the rest frame of fluid element) $\epsilon_{\rm sw}=0.5$~GeV/fm$^3$. Mean spin vector of produced $\Lambda$, as well as other baryons with spin $S$ and mass $m$, is calculated at this hypersurface with the help of a Cooper-Frye-like formula \cite{Becattini:2013fla}:
\begin{equation}\label{eq-Pi}
  S^\mu =\frac{1}{N} \frac{(-1)}{2m}\frac{S(S+1)}{3} 
  \int\frac{d^3 p}{p^0} \int d\Sigma_\lambda p^\lambda f(x,p) (1-f(x,p)) \epsilon^{\mu\nu\rho\sigma} p_\sigma \varpi_{\nu\rho},
\end{equation}
where $N=\int\frac{d^3p}{p^0}\int d\Sigma_\lambda p^\lambda f(x,p)$ is the average number of 
given sort of hadrons produced at the particlization surface. Hadrons produced at the particlization surface we call below as primary hadrons. Finally, the polarization vector is obtained by normalizing the mean spin vector by the spin of the particle, $P^\mu=S^\mu/S$.

UrQMD cascade does not include polarization degrees of freedom. Therefore, unlike the basic studies performed with this model \cite{Karpenko:2015xea}, UrQMD cascade is not used for the post-hydro stage in this work.

\section{Results and discussion}\label{sec2}

\begin{figure}[h]
\includegraphics[width=0.52\textwidth]{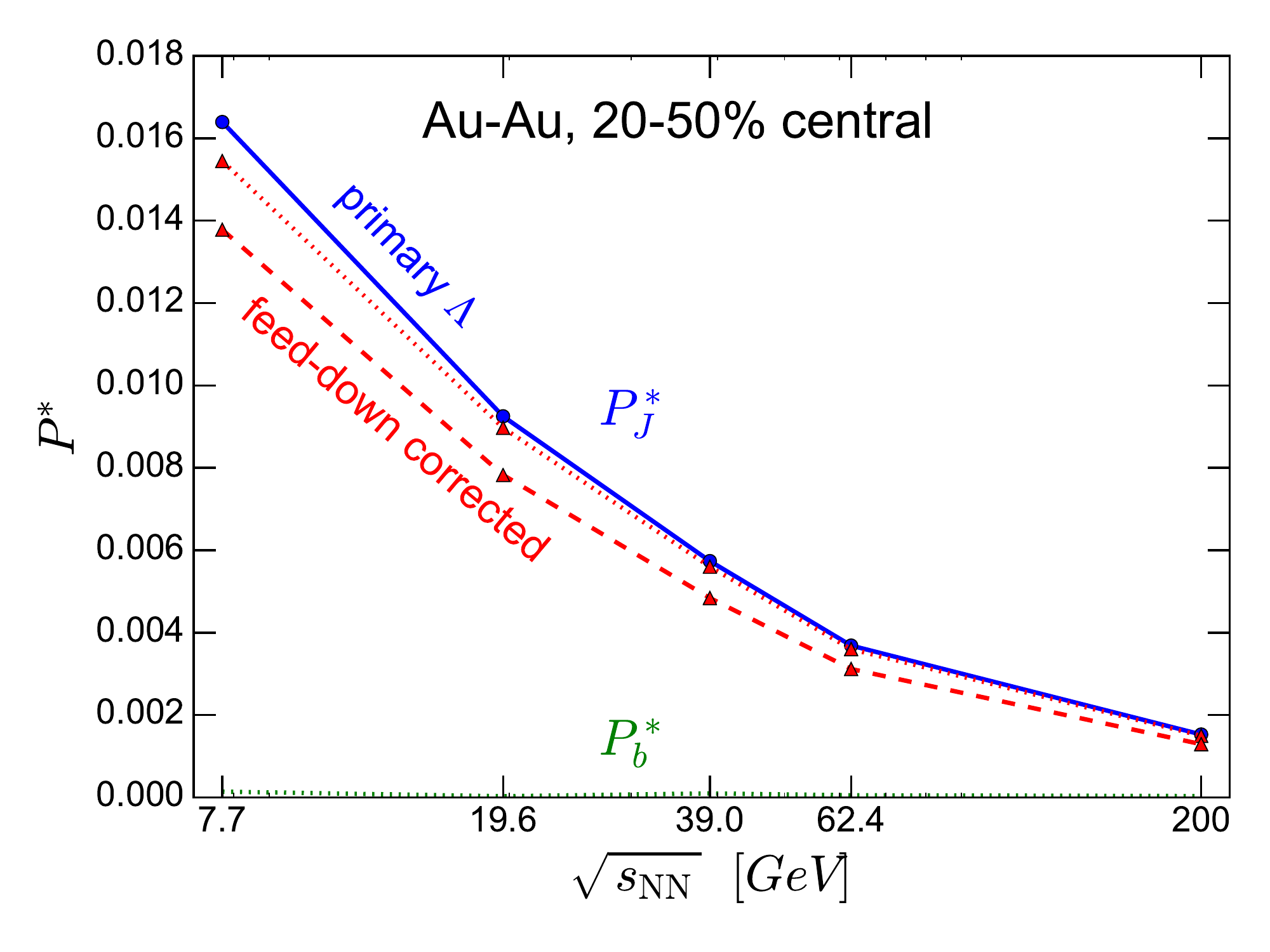}
\includegraphics[width=0.52\textwidth]{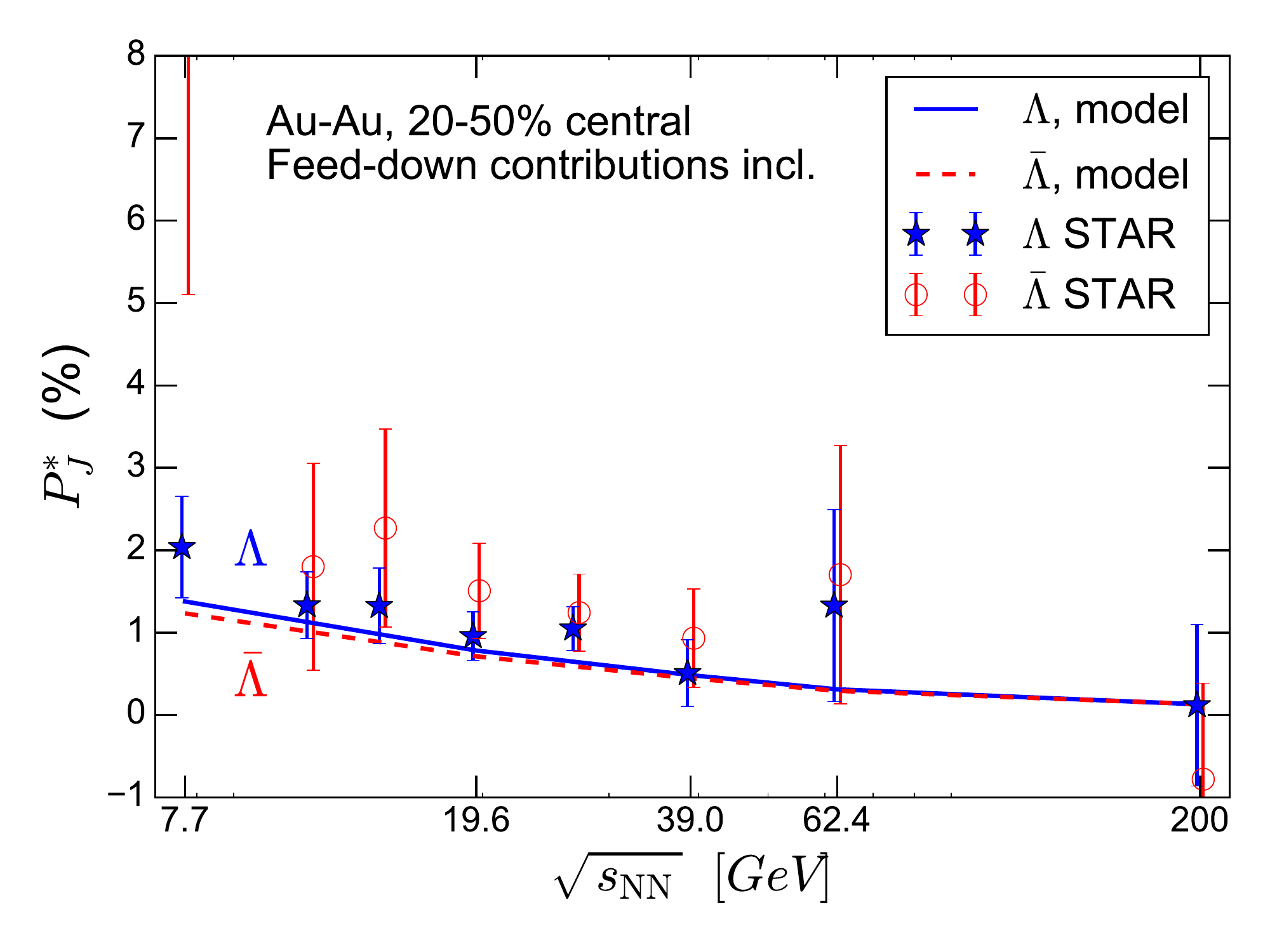}
\vspace*{-20pt}
\caption{Left panel: collision energy dependence of mean polarization of primary $\Lambda$ (solid curve), with feed-down contributions from $\Sigma(1385)$ and $\Sigma^0$ (dotted curve) and with feed-down corrections from resonances up to $\Sigma(1670)$ (dashed curve). Right panel: collision energy dependence of mean polarization of $\Lambda$ and $\bar\Lambda$ including feed-down corrections from resonances up to $\Sigma(1670)$ (and respective antiparticles) in comparison with STAR measurements \cite{STAR:2017ckg}.}\label{fig1}
\end{figure}

Parameters of the model are adjusted for optimal reproduction of experimental data for rapidity distributions (NA49 data for Pb-Pb collisions at SPS energies and PHOBOS data for certain collision energies in the BES range), transverse momentum spectra of identified hadrons and momentum integrated elliptic flow, reported in \cite{Karpenko:2015xea}. To evaluate the mean polarization of $\Lambda$ hyperons, we run the model with exactly the same setup for 20-50\% central Au-Au collisions at RHIC BES energies. For the mean polarization vector, the only non-zero component is $P_J$, which is directed along the angular momentum of the system, i.e.\ perpendicular to the reaction plane. This component is shown on Fig.~\ref{fig1} as a function of collision energy.

{\bf Collision energy dependence.} First of all, one can observe that the mean polarization of primary $\Lambda$ (i.e.\ the ones produced at particlization hypersurface) significantly decreases with increasing collision energy, from top value of about 1.6\% at $\sqrt\sNN=7.7$~GeV to less than 0.2\% at full RHIC energy. To explain such decrease, let us first discuss how the polarization emerges in the model. From Eq.~\ref{eq-Pi}, one can see that polarization (in the limit where it is small) is linearly related to the local vorticity, integrated with local particle current over the hypersurface. We have found that for the $P_J$ component the dominant contribution comes from the term proportional to $\varpi_{xz}$ component of vorticity. Furthermore, as the yield of $\Lambda$ is dominated by low-$p_T$ contributions, which are produced from the fluid at late stages of its evolution, the mean polarization is largely determined by $\varpi_{xz}$ at the late stages (late times) of the hydro phase.

\begin{figure}[h]
\includegraphics[width=0.52\textwidth]{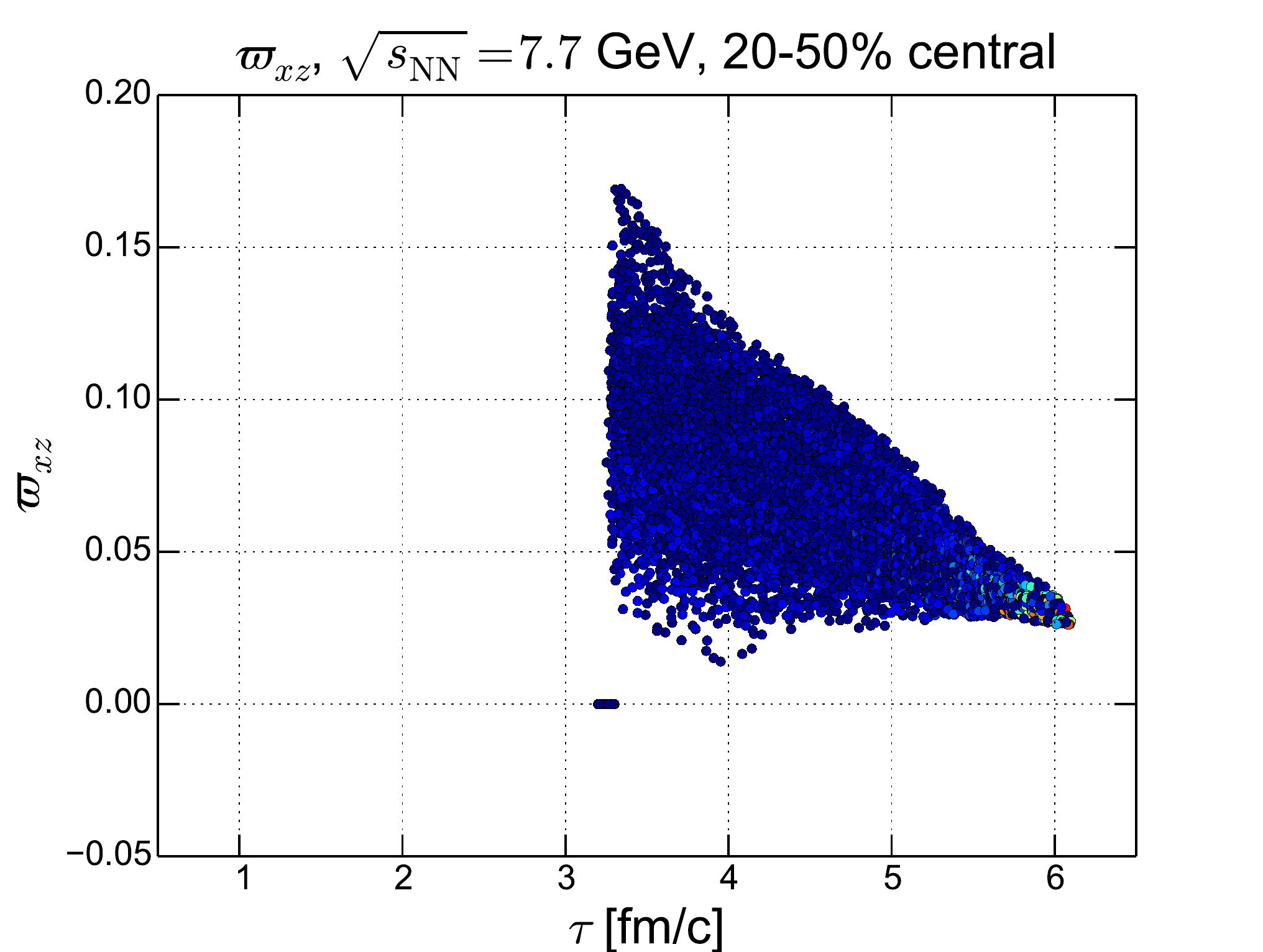}
\includegraphics[width=0.52\textwidth]{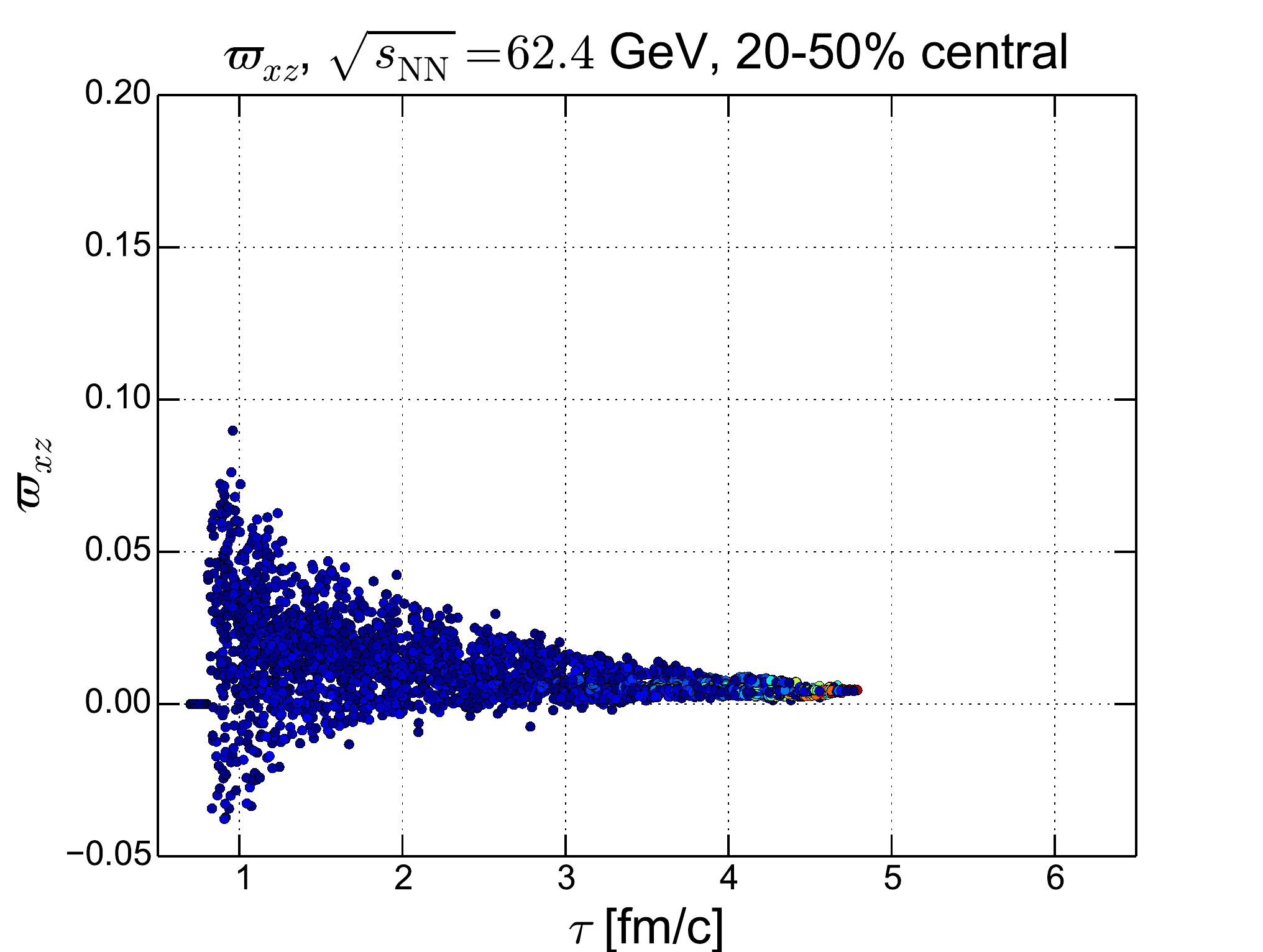}
\vspace*{-20pt}
\caption{Time distributions of $\varpi_{xz}$ component of thermal vorticity over the particlization hypersurface, visualized from two hydrodynamic calculations with averaged initial state corresponding to 20-50\% central Au-Au collisions at $\sqrt\sNN=7.7$~GeV (left panel) and $\sqrt\sNN=62.4$~GeV (right panel).}\label{fig2}
\end{figure}

We visualize the time distribution of $\varpi_{xz}$ over the particlization hypersurface from two hydro calculations with averaged initial state corresponding to $\sqrt\sNN=7.7$ and $62.4$~GeV in Fig.~\ref{fig2}. One can indeed see that at late times the $\varpi_{xz}$ is significantly smaller at the higher collision energy than at the lower one. It is a result of two effects: 1) different initial distribution of $\varpi_{xz}$ at high and low collision energies: at lower collision energies, baryon stopping creates a shear flow profile around midrapidity resulting in large same sign $\varpi_{xz}$, whereas at higher energies because of approximate boost invariance at midrapidity the initial $\varpi_{xz}$ is smaller and does not have definite sign 2) longer lifetime of fluid stage at high energies, which dilutes the initial vorticity stronger.

We would like to point out that qualitatively similar collision energy dependence of mean $\Lambda$ polarization has been found in a framework of ideal hydro model with non-boost-invariant Yang-Mills flux-tube initial state, applied to RHIC BES energies \cite{Xie:2017upb}.

{\bf Feed-down contributions.} Primary $\Lambda$ are only about 25-28\% of all $\Lambda$ produced. The rest is feed-down contributions from decays of cocktail of heavier hyperons, which are themselves polarized via the same mechanism. The largest sources of feed-down contributions are $\Sigma^*$ and $\Sigma^0$ decays.

It can be shown that for the transfer of mean, momentum-integrated, spin vector in the rest frame of particle, a linear rule applies \cite{Becattini:2016gvu}:
\begin{equation}{\bf S}^*_D = C {\bf S}^*_X,\label{eq-transfer}\end{equation}
where ${\bf S}$ denotes the mean spin vector, D is daughter particle, X is a parent resonance and coefficient $C$ is calculated for every particular decay process. For strong and electromagnetic decays, which conserve parity, the coefficients $C$ turns out to be independent on dynamical decay amplitudes, and is determined by Clebsch-Gordan coefficients and values of spin and parity of the resonance and its decay products. This coefficient has been calculated for various resonance decays and the values are reported in \cite{Becattini:2016gvu}.

Eq.~\ref{eq-transfer} allows for a straightforward estimate of feed-down corrected polarization, accoridng to :
\begin{eqnarray}\label{eq-res-correction}
  {\bf S_\Lambda}^*= \frac{N_\Lambda + \sum\limits_{X}N_X\left[ C_{X\rightarrow\Lambda}b_{X\rightarrow\Lambda} -
 \frac{1}{3} C_{X\rightarrow\Sigma^0}b_{X\rightarrow\Sigma^0}\right]}
 {N_\Lambda + \sum\limits_{X}b_{X\rightarrow\Lambda}N_{X} + \sum\limits_Xb_{X\rightarrow\Sigma^0}N_X}
 \cdot {\bf S}^*_{\Lambda,{\rm prim}},
\end{eqnarray}
where the sum goes over resonances $X$ which have $\Lambda$ or $\Sigma^0$ as a decay product, $C_{X\rightarrow\Lambda}$ and $C_{X\rightarrow\Sigma^0}$ 
are polarization transfer coefficients, $b_{X\rightarrow\Lambda}$ and $b_{X\rightarrow\Sigma^0}$ are 
the branching ratios for decay channels yielding in $\Lambda$ and $\Sigma^0$ respectively. Here we use the fact that most of $\Sigma^0$ decay to $\Lambda \gamma$ with polarization transfer $C=-1/3$, which allows to treat cascade decay contributions $X\rightarrow\Sigma^0\rightarrow\Lambda$.

The polarization of feed-down corrected $\Lambda$ is shown on Fig.~\ref{fig1}. It is interesting to note that when we only include feed-down corrections from $\Sigma(1385)$ and $\Sigma^0$, the resulting polarization, represented by dotted line on the left panel, is almost unchanged because of interplay of larger polarization of primary $\Sigma(1385)$ due to their spin $3/2$, and particular values of polarization transfer coefficients. However, as we take more resonances into account, up to $\Sigma(1670)$, the additional effect is essentially 15\% ``dilution'' of the resulting polarization, represented by the dashed line on the same plot.

In addition to that, $\Lambda$ actively rescatter in hadronic phase, as they have cross-sections comparable to those of nucleons. Hadronic elastic interaction may involve a spin flip which, presumably, will randomize the spin direction of primary as well as secondary particles, thus decreasing the estimated {\em mean} polarization. We leave the estimate of hadronic rescattering effect on polarization for future studies.

{\bf $\Lambda$ and $\bar\Lambda$.} Finally, we have calculated the polarization of anti-$\Lambda$ hyperons, which is shown on the right panel of Fig.~\ref{fig1}. Whereas at lower energies the yield of $\bar\Lambda$ in the model is considerably smaller than $\Lambda$ due to baryon chemical potential, their polarizations are very close. Although theory curve for $\Lambda$ agrees with experimental data within error bars at all collision energies, the splitting between $\Lambda$ and $\bar\Lambda$ is much smaller than the trend from experimental points, and has opposite sign. Another possible effect which can create the splitting, and is not included in our calculations, is coupling of magnetic moment to magnetic field at hadronization, which aligns spins of $\Lambda$ and $\bar\Lambda$ in opposite directions because of opposite magnetic moments of $\Lambda$ and $\bar\Lambda$.

{\bf Acknowledgements.} This work was partly supported by the University of Florence grant {\em Fisica dei plasmi relativistici: teoria e applicazioni moderne}. The simulations have been performed at the Center for Scientific Computing (CSC) at the Goethe-University Frankfurt.


\bibliographystyle{elsarticle-num}
\bibliography{mybib}







\end{document}